\documentclass[apj]{emulateapj-rtx4}

\newcommand{\gtres}{G331.5$-$0.1}
\newcommand{\gtresoutflow}{G331.512$-$0.103}
\newcommand{\kms}{km s$^{-1}$}
\newcommand{\lsun}{\mbox{L$_\odot$}}
\newcommand{\msun}{\mbox{M$_\odot$}}
\newcommand{\ee}[1]{\mbox{${} \times 10^{#1}$}}
\newcommand{\eten}[1]{\mbox{$10^{#1}$}}
\newcommand{\jj}[2]{\mbox{$J = #1\rightarrow#2$}}
\newcommand\cmv{\mbox{cm$^{-3}$}}
\newcommand\cmc{\mbox{cm$^{-2}$}}
\newcommand{\hcop}{\mbox{HCO$^+$}}
\newcommand{\hcopt}{\mbox{H$^{13}$CO$^+$}}

\newcommand{\tcross}{\mbox{$t_{cross}$}} 
\newcommand{\vmax}{\mbox{$v_{max}$}} 

\shorttitle{ALMA observations of \gtres}
\shortauthors{Merello et al.}

\begin{document}

\title{ALMA observations of the massive molecular outflow \gtresoutflow}

\author{Manuel Merello$^{1,2}$, Leonardo Bronfman$^1$, Guido Garay$^1$, Nadia
Lo$^1$, Neal J. Evans II$^2$, Lars-\AA ke Nyman$^3$, \\Juan R. Cort\'es$^3$, Maria R. Cunningham$^4$}

\affil{$^1$Departamento de Astronom\'ia, Universidad de Chile, Casilla 36-D, Santiago, Chile}
\affil{$^2$The University of Texas at Austin, Department of Astronomy, 2515 Speedway, Stop C1400, Austin, Texas 78712-1205}
\affil{$^3$Joint ALMA Observatory (JAO), Alonso de Cordova 3107, Vitacura,
Santiago, Chile}
\affil{$^4$School of Physics, University of New South Wales, Sydney, NSW 2052,
Australia}

\begin{abstract}
The object of this study is one of the most energetic and luminous molecular outflows known in the Galaxy, \gtresoutflow. Observations with ALMA Band 7 (350 GHz; 0.86 mm) reveal a very compact, extremely young bipolar outflow and a more symmetric outflowing shocked shell surrounding a very small region of ionized gas. The velocities of the bipolar outflow are about 70 \kms\ on either side of the systemic velocity. The expansion velocity of the shocked shell is $\sim$24 \kms, implying a crossing time of about 2000 yr. Along the symmetry axis of the outflow, there is a velocity feature, which could be a molecular ``bullet'' of high-velocity dense material. The source is one of the youngest examples of massive molecular outflow found associated with a high-mass star.

\end{abstract}


\section{Introduction}

The formation of massive stars has a major impact on the surrounding
interstellar medium, and regions of massive star formation dominate
the star formation appearance of galaxies. While during the last couple of decades significant progress (both observational and theoretical) has been made in the study of massive stars and their parental molecular clouds, we are still far away from a complete picture of their formation process \citep[for a recent discussion about this subject, see][]{kennicutt2012}

Massive stars rapidly affect their surroundings, so catching
one early in the formation process can provide valuable insights. Since
they form rarely compared with their low-mass counterpart, most are at large distances, making high resolution
observations challenging. The Atacama Large Millimeter/submillimeter Array (ALMA) provides the capability to study distant
objects in the Galaxy with good spatial resolution and sensitivity.

The G331.5-0.1 molecular cloud is a giant, elongated (178 by 41 pc) cloud in
the Norma spiral arm, at a distance of 7.5 kpc, with a total CO mass of $8.7\ee6$ \msun\ (M$_{\mathrm{LTE}}=3.5\ee6$ \msun\ from C$^{18}$O emission) and a total infrared luminosity of 3.6\ee6 \lsun\
(\citealt{Bronfman2008, Merello2013}),
making it one of the most active and extreme high-mass
star forming environments in our Galaxy.
The central region hosts an extended \ion{H}{2} region/infrared source and
a number of ultracompact \ion{H}{2} (UCHII) regions. Maps at submillimeter
wavelengths show six continuum sources, with masses above \eten3 \msun\
and average gas surface densities of 0.4 g \cmc. 
Single dish spectral line observations with the Atacama Submillimeter Telescope Experiment and the Atacama Pathfinder EXperiment (APEX) of one of these sources, MM3, revealed extremely broad ($\pm 80$ \kms)
wings on lines of CO, CS (Bronfman et al. 2008), SO, and SiO
(Merello et al. 2013). This highly massive and energetic outflow, \gtresoutflow, is unresolved spatially in a
7\arcsec\ beam.

Here we present observations with ALMA that resolve the \gtresoutflow\ molecular outflow,
allowing derivation of the outflow properties including age. In forthcoming papers we will describe the full set of molecular lines observed with ALMA Band 7.

\section{Observations and results}

The observations were performed on 2012 January 12 at frequencies around
350 GHz (0.86 mm) with ALMA Band 7 in Cycle 0 with 
17 (12-m diameter) antennas in the compact configuration, with baselines 
ranging from 18.5 m to 269 m. We observed a single field of view centered at $\alpha_{2000} = 16^h12^m10.09^s$, $\delta_{2000} = -51\arcdeg 28\arcmin 38.4\arcsec$. The primary beam was 17.8\arcsec\ and the synthesized beam was $1.38\arcsec \times 0.68\arcsec$, 
with a position angle (PA) of $-37.6$\arcdeg.
The integration time was 34 minutes on source, with an additional 24
minutes on calibrators. The nearby quasar J1604-446 was used for flux calibration, and J1427-421, located at 27.7\arcdeg\ from the \gtresoutflow\ source, was used for phase and bandpass calibrations. The data were processed using the Common Astronomy Software Application \citep[CASA;][]{mcmullin2007}, and the resulting maps have cells with sizes of 0.14\arcsec, with 254 cells in each spatial dimension. Our images were obtained considering a ``briggs" weighting mode on the data with robust=0.5.
The four spectral windows have a bandwidth of 1875 MHz containing 3840
channels. The separation between each channel is 488.281 kHz (0.4215 km s$^{-1}$
at 357.3 GHz). The bands were centered at 345.79599 GHz, 347.2 GHz, 357.3 GHz and 358.60587 GHz.

While a rich spectrum of lines was observed, this Letter focuses on the lines of SiO(8-7) (347.33058 GHz), H$^{13}$CO$^+$(4-3) (346.99835 GHz), HCO$^+$(4-3) (356.73424 GHz), and CO(3-2) (345.79599 GHz), 
mostly diagnosing the outflow and its interaction with the
surrounding cloud.  Table 1 presents the four lines reported in the present 
work, the noise for each, the peak flux density, and the velocity at the peak.
The rms noise was assessed over a region of 35\arcsec$\times$35\arcsec\ 
centered on the source, avoiding a square area of 11\arcsec\ in size 
centered on the emission.
For line-free channels, the noise was $\sim$0.014 Jy beam$^{-1}$. However, 
for strong emission lines, the noise is larger, presumably because
of incomplete $uv$ coverage. On the channels of maximum intensity, 
toward the systemic velocity of the source, the noise increases substantially. 
To be conservative, we use the noise values at the line velocities
in setting thresholds for contours, etc. Moment 0 maps were obtained for each emission line over its full velocity extension, and FWHM angular sizes and P.A.s of these maps were fitted using the task IMFIT of CASA (Columns 6 and 7 of Table 1). Physical sizes, considering a source distance of 7.5 kpc, are presented in Column 8 of Table 1.

Figure 1 shows the integrated spectra for the lines described in Table 1. The integration is made over a box of 10.5\arcsec$\times$10.5\arcsec, centered at $\alpha_{2000} =
16^h12^m09.965^s$, $\delta_{2000} = -51\arcdeg28\arcmin 38.250\arcsec$. Very wide wings (up to $\pm 70$ \kms) are seen on the SiO \jj87, CO
\jj32, and \hcop\ \jj43\ lines.
The \hcopt\ \jj43\ line is narrow toward the systemic velocity of the source ($-88.9$ \kms), but also shows traces of emission over the red wing of the outflow.
The high velocity wings in the SiO profile match those in the APEX spectra presented in
\cite{Merello2013}, with differences between both profiles of less than 10\%, 
indicating that the ALMA observations recover
all the emission seen in the single-dish spectra.
The CO spectrum presents several absorption features at different velocities.
Besides the self-absorption component at the systemic velocity of the source, an absorption dip at $\sim-100$ \kms\ is related to a second
velocity component in the central region of the \gtres\ giant molecular cloud, while the large feature
between $-70$ and $-52$ \kms\ and the one at $-40$ \kms\ are related 
to foreground galactic emission identified in \cite{Bronfman89}. 
The self-absorption dip at $-88.9$ \kms\ is also present in the \hcop\ profile and
weakly in  the \hcopt\ profile. 

The \hcopt\ profile resembles a typical blue-profile found in regions 
with infalling material \citep{myers2000}, but observations of a more
optically thin molecule, like HC$^{18}$O$^+$, will be needed to test this 
hypothesis. The \hcopt\
profile also shows a trace of emission in the red wing of the outflow,
with a secondary peak of emission at $\sim -50$ \kms. 

Figure 2 shows the channel maps of the SiO line emission. The crosses
indicate the peak positions at velocities about $\pm 40$ \kms\ from 
the systemic velocity. These channels define a
``bipolar axis" of symmetry of $102.5\pm 1.6$\arcdeg. 
Experiments using other channels, different signal-to-noise levels, etc.
resulted in variations around this value of only 1.6\arcdeg, which we assign
as the uncertainty.
This angle differs slightly from the P.A. found in the deconvolved moment 0 map of SiO, (P.A. of 105\arcdeg; see Table 1).

The integrated intensity SiO map shows a ring-type feature toward the systemic velocity of the source ($-90\pm15$ \kms). We used the 3 $\sigma$ contour level to estimate the outer edge of this emission, obtaining major and minor axes of  5.12$\pm0.16$\arcsec\ and 4.52$\pm0.33$\arcsec, respectively, with a P.A. of 140.6\arcdeg. The deconvolved sizes are then  4.79$\pm$0.19\arcsec\ and 4.42$\pm$0.35\arcsec\ (0.17 pc at the source distance), with a P.A. of 137\arcdeg.

The P.A. of the beam elongation is 175\arcdeg\ from this angle, so
some of the extension may be beam-related, but the extent is considerably
larger than the beam, and the deconvolution should remove that effect.
The inner cavity is centered at $\alpha_{2000} =
16^h12^m10.00^s$, $\delta_{2000} = -51\arcdeg28\arcmin 37.45\arcsec$, it has a size of 0.99\arcsec$\times$0.66\arcsec, with a PA of
160\arcdeg, and it is determined by the 16$\sigma$ emission contour. Therefore, the cavity in the SiO emission at the
systemic velocity is not resolved with the present beam size. 
The SiO map also shows clumpy, irregular structure at the systemic velocity, with two peaks of emission of 5.0 and 3.9 Jy beam$^{-1}$, located symmetrically at both sides of the inner cavity.

Figure 3 shows the intensity map of \hcopt\ in color, overlaid with SiO
contours in black.
The line maps are at $\sim-91.9$ \kms, near the systemic velocity of the 
source, where the \hcopt\ profile shows the peak of intensity. 
The \hcopt\ emission shows a ring shape similar to that seen in SiO, with a inner hole coincident with the cavity found in the SiO observations, 
but with most of this emission lying outside the SiO ring. The \hcopt\ inner hole has a size of 1.82\arcsec$\times$1.12\arcsec, with a PA of 112\arcdeg, and it is determined by the 4 $\sigma$ emission contour. The \hcopt\ structure observed at this velocity channel 
has major and minor axes, considering the 3 $\sigma$ contour as the outer edge of this emission, of 8.10$\pm$0.32\arcsec\ and 6.79$\pm$0.20\arcsec, respectively, with a PA of 78\arcdeg. The deconvolved angular sizes are then
8.03$\pm$0.32\arcsec\ and 6.57$\pm$0.22\arcsec, with a PA of
76\arcdeg, which correspond to a geometric mean size of 0.26 pc at the source distance.

The red ellipse in Figure 3 represents the 50\% contour of the 
radio continuum source at 8.6 GHz 
(Merello et al. 2013). 
The emission at 8.6 GHz is located inside the inner cavity of
the SiO emission and is unresolved within the beam 
(1.51\arcsec$\times$0.99\arcsec).

\section{Discussion}

The SiO emission is an indicator of shocked gas,
commonly seen around young stellar objects,
with enhanced abundance attributed to Si being sputtered from dust grains
for jet velocities exceeding 25 \kms\
\citep[e.g.,][]{arce2007, garay98, garay02}.
Because the wing velocities of the \gtresoutflow\ outflow extend to
$\pm 70$ \kms, we expected to see SiO emission in the wings, likely
with a bipolar pattern, and we do. However, the fact
that a ring of  SiO emission is also present at the systemic velocity,
with a cavity of the size of the beam, indicates that a more isotropic
high speed wind is producing shocked gas in a shell. We interpret then that the \gtresoutflow\ source corresponds to a jet almost in the line of sight with an expanding shocked shell surrounding its driving source, which in projection is observed as a ring at the systemic velocity. 

To explore this hypothesis, we construct position-velocity (PV) 
diagrams of the SiO and \hcopt\ emission (Figure 4), 
for slices along the bipolar axis and perpendicular to it.
The cavity is clearly present in all diagrams, centered near $v = -90$ \kms. 
The SiO PV diagram along the secondary axis shows two strong peaks nearly
symmetric around the cavity. These two peaks extend around the cavity down to the 16 $\sigma$ emission contour (represented in red in Figure 4), extending in velocity between $-111.9$ and $-63.8$ \kms. The PV diagram along the bipolar axis shows a strong peak at an angular
offset of 1\arcsec.
The 3 $\sigma$ contour of the \hcopt\ traces the red wing of the outflow 
up to a velocity of $\sim -25$ \kms.
An additional peak is observed at
$-45$ \kms\ along the primary axis, but at $-50$ \kms\ along the secondary axis. This emission extends between $-59<v_{lsr}<-29$ \kms, defined by the 6 $\sigma$ contour in the \hcopt\ PV diagram along the symmetry axis, and it has a size of 2.5\arcsec$\times$1.4\arcsec ($\sim0.07$ pc). We consider this high-velocity emission likely to be a molecular ``bullet'', feature sometimes observed in star formation regions  \citep[e.g.][]{masson93, santiago2009, kristensen2012}.  

The bipolar symmetry at the extreme velocities requires a jet-like outflow, 
but the ring structure at the systemic velocity indicates a nearly isotropic
component. The latter surrounds a region of ionized material, whose
expansion could produce the isotropic component. 
Considering the extension of the symmetric peaks in the PV diagram of SiO, we estimate that
the expanding shell
has a velocity of $\sim$24 \kms. A jet originating from the high-mass forming star,
the likely  source of the radio emission observed at 8.6 GHz, could
drive the molecular outflow observed in the high-velocity wings of the SiO
spectra. The \hcopt\ emission, being optically thin toward the wings of the
outflow, is only tracing the red lobe of this outflow, and the line profile
suggests that toward the systemic velocity, some gas is still infalling
toward the central star. However, such a dip can also be created by resolving
out ambient cloud emission, so further observations would be needed to 
test this interpretation.

An estimate for the crossing time is given by 
$ \tcross  = \theta/(2 \vmax)$, 
with $\theta$ the diameter of a projected shell, and \vmax\ the maximum
velocity relative to the ambient gas. 
Using $\vmax = 24$ \kms, we obtain
$\tcross < 600$ yr for the inner hole and $\sim$2000 yr for the SiO
ring with diameter 3\arcsec\ (or 0.11 pc, considered as the middle point between the inner and outer edges of the ring).
These estimates would make this source
one of the youngest examples of flows around massive forming stars.
The momentum ($2.4\ee3$ \msun\ \kms) and kinetic energy (1.4\ee{48} erg)
of the outflow were measured by \citet{Bronfman2008} from CO(7-6) observations. 
Taking 2000 years as the relevant time,
the resulting force is 1.2 \msun\ \kms\ yr$^{-1}$ and the mechanical 
luminosity is 6.4\ee3 \lsun, both higher than any values found in the
compilation of outflow properties by \citet{wu04}.

The mass of the \hcopt\ emission toward the ambient gas and the bullet is estimated considering a local thermodynamic equilibrium (LTE) formalism and optically thin emission \citep[see e.g.][]{garden1991}. Considering an excitation temperature of 100 K, a mean molecular weight $\mu$ = 2.72 $m_{\mathrm H}$, and an abundance [\hcopt/H$_2$] = 3.3$\times10^{-11}$ \citep{blake1987}, the LTE mass of the ambient gas and the bullet are 36 \msun\ and 1.8 \msun, respectively.

The continuum emission was determined from regions of the ALMA Band 7 spectrum that
appeared to be line free, though we cannot rule out some level of 
contamination by line emission. 
The lower panel of Figure~\ref{fig:h13cop_sio} 
shows the 0.86 mm continuum emission, overlaid on the H$^{13}$CO$^+$ emission at $-91.9$ km s$^{-1}$.
A point-like source lies within the ring of SiO and
\hcopt\ emission, and a more extended component agrees well with the \hcopt\
ring.

The continuum emission has a
FWHM of 2.94\arcsec$\times$2.18\arcsec\ and a total flux of 4.3 Jy. This emission could arise from either free$-$free emission from a very dense
ionized wind or from dust emission, or a combination.
As a test, we estimated the integrated flux of the continuum emission with a point source removed
at the peak position of the 8.6 GHz continuum emission showed in Figure 3, obtaining a value of 3.9 Jy. 
This integrated flux of the continuum emission is much less than the 66.7 Jy measured at this wavelength in a
33\arcsec\ region \citep{Merello2013}, but much more than would be expected from a uniformly distributed emission, therefore we associate this emission with an envelope around the central object. If we attribute this measured flux to dust emission, we can estimate mass and other properties
of the source. At a radius of 1\arcsec, or 0.036 pc at the source distance, and a luminosity
of 7\ee5 \lsun\ \citep{Merello2013}, the equilibrium dust temperature
would be $400-1000$ K, for a range of dust properties.
Assuming 400 K, and using OH5 opacities, the mass estimated in a region of 
2.9\arcsec\ in size, or 0.11 pc, would be 38 \msun, implying a mass
surface density of 5140 \msun\ pc$^{-2}$, or 1.07 g cm$^{-2}$, and a mean
volume density of 1.3\ee6 \cmv.

An interesting comparison can be made to the outflow from G5.89
(e.g., \citealt{harvey88, acord97, watson07, su12}), which also
shows a shell morphology along with a bipolar flow. The expansion
age for the UCHII region is 600 yr, measured from multi-epoch observations
\citep{acord98}, while the bipolar outflow is older, with estimates
ranging from 3000 to 7700 yr. Estimates of the outflow properties
vary, but the values in the \citet{wu04} compilation are a
mass of 70 \msun, a momentum of 1600 \msun \kms, and a  mechanical
luminosity of 1600 \lsun.
In comparison with the G5.89 source, the \gtresoutflow\ outflow has a shorter age and a higher mechanical luminosity and momentum.

\section{Summary and conclusions}

We have performed a study with ALMA Band 7 of the massive and energetic molecular outflow \gtresoutflow. The high angular resolution observations allow us to resolve this source in a set of four molecular lines: SiO(8-7), H$^{13}$CO$^+$(4-3), HCO$^+$(4-3) and CO(3-2). The SiO emission is confined in a region of size less than 5\arcsec\ (0.18 pc at a distance of 7.5 kpc), and reveal the presence of a ring-type structure toward the systemic velocity of the source. This feature is also observed in the \hcopt\ line, and the cavity is coincident with a strong and compact radio continuum source observed at 8.6 GHz. We interpret these observations as a young stellar object producing a compact \ion{H}{2} region, with an expansion shock propagating into the medium and possibly with dense material still infalling around this shell. CO, HCO$^+$ and SiO observations trace the full extension of the outflow wings, but the H$^{13}$CO$^+$ only traces the red wing, also showing the possible presence of a bullet of dense material. We estimated an age of $\sim$2000 yrs for the expanding shell, making it one of the youngest examples of massive molecular outflows observed toward massive young stars.

\acknowledgements{M.M. and N.J.E. gratefully acknowledge support from the NSF grant AST-1109116. L.B. and G.G. acknowledge support from CONICYT project BASAL PFB-06. N.L.'s postdoctoral fellowship is supported by CONICYT/FONDECYT postdoctorado, under project no. 3130540. N.L. acknowledges support from the ALMA-CONICYT Fund for the Development of Chilean Astronomy Project 31090013. This Letter makes use of the following ALMA data: ADS/JAO.ALMA\#2011.0.00524.S. ALMA is a partnership of ESO (representing its member states), NSF (USA) and NINS (Japan), together with NRC (Canada) and NSC and ASIAA (Taiwan), in cooperation with the Republic of Chile. The Joint ALMA Observatory is operated by ESO, AUI/NRAO and NAOJ.}

\clearpage
\begin{deluxetable}{lccccccc}
\tabletypesize{\scriptsize}
\tablecaption{Parameters of the molecular line observations\label{tbl-obs}}
\tablewidth{0pt}
\tablehead{
\colhead{Line} & \colhead{Frequency} & \colhead{Noise} & \colhead{Velocity at
peak} & \colhead{Peak flux density}&\colhead{Deconv. angular size\tablenotemark{a}}&\colhead{Position angle}&\colhead{Deconv. size\tablenotemark{a}}\\
&\colhead{(GHz)} & \colhead{(Jy beam$^{-1}$)}& \colhead{(\kms)} &
\colhead{(Jy)}&\colhead{(\arcsec)}&\colhead{(\arcdeg)}&\colhead{(pc)}
}
\startdata

SiO(8-7) & 347.33058 & 0.139 & $-89.5311$ & 40.4 & 2.6$\times$2.3 & 105 & 0.09\cr
H$^{13}$CO$^+$(4-3) & 346.99835 & 0.045 & $-91.9158$ & 30.1 & 3.5$\times$2.4 & 93 &0.11\cr
HCO$^+$(4-3) & 356.73424 & 0.214 & $-85.0056$ &91.4 & 2.6$\times$2.5 & 106&0.09\cr
CO(3-2) & 345.79599 & 0.365 & $-96.9012$ & 238.5 & 2.8$\times$2.5 & 110&0.10

\enddata
\tablenotetext{a}{FWHM of moment 0 maps, integrated over full velocity extension of the line.}
\label{tbl:observ_mol}
\end{deluxetable}

\clearpage
\begin{figure}[h]
	\begin{center}
		\includegraphics[width=0.6\textwidth]{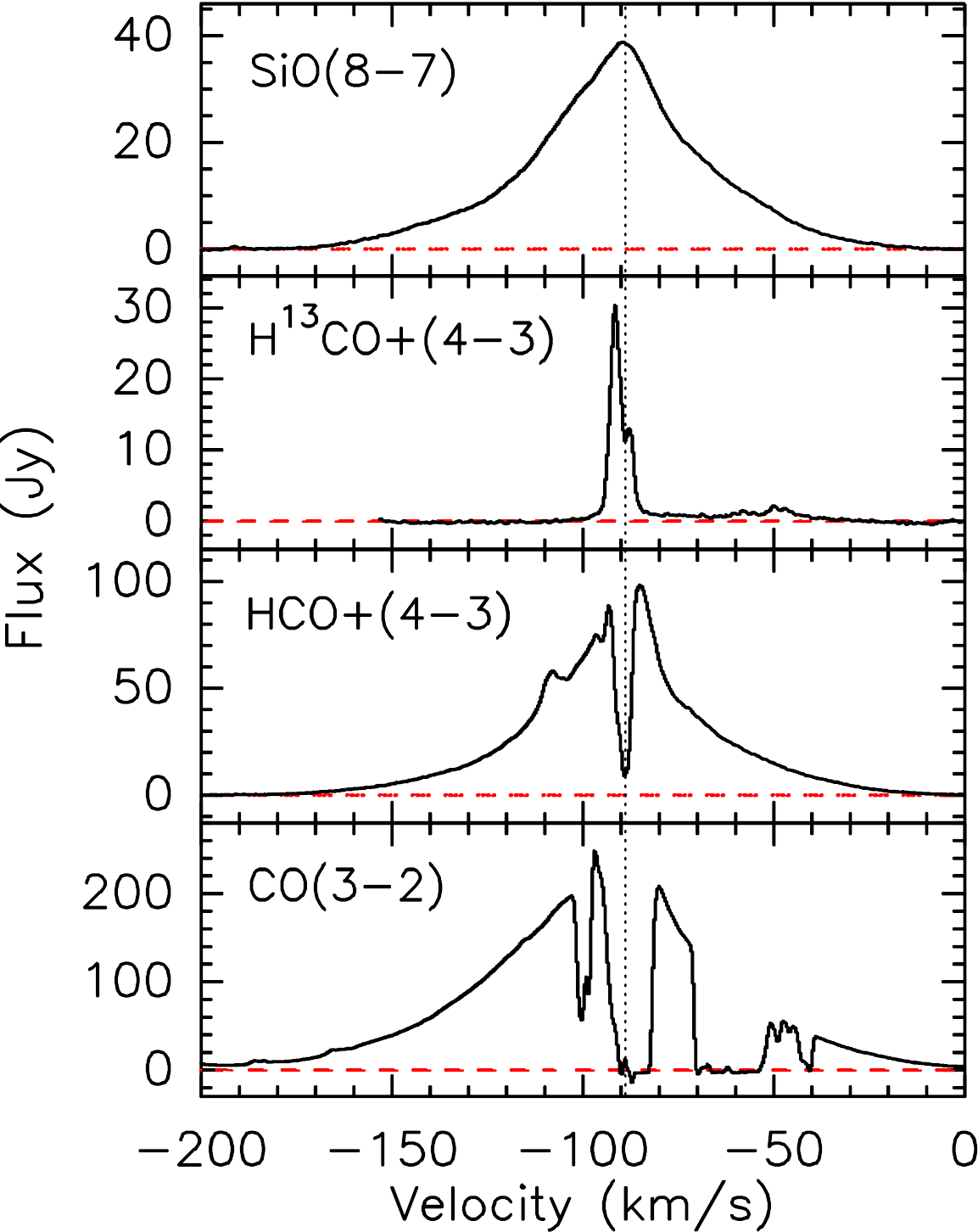}
	\end{center}
	\caption{Integrated spectra observed with ALMA of the \gtresoutflow\ molecular
outflow. A dashed line is drawn in each spectrum representing the baseline. The dotted vertical line shows the systemic velocity of this source ($-88.9$ \kms).}
	\label{fig:spectra}
\end{figure}

\begin{figure}[h]
	\begin{center}
		\includegraphics[angle=0, width=1\textwidth]{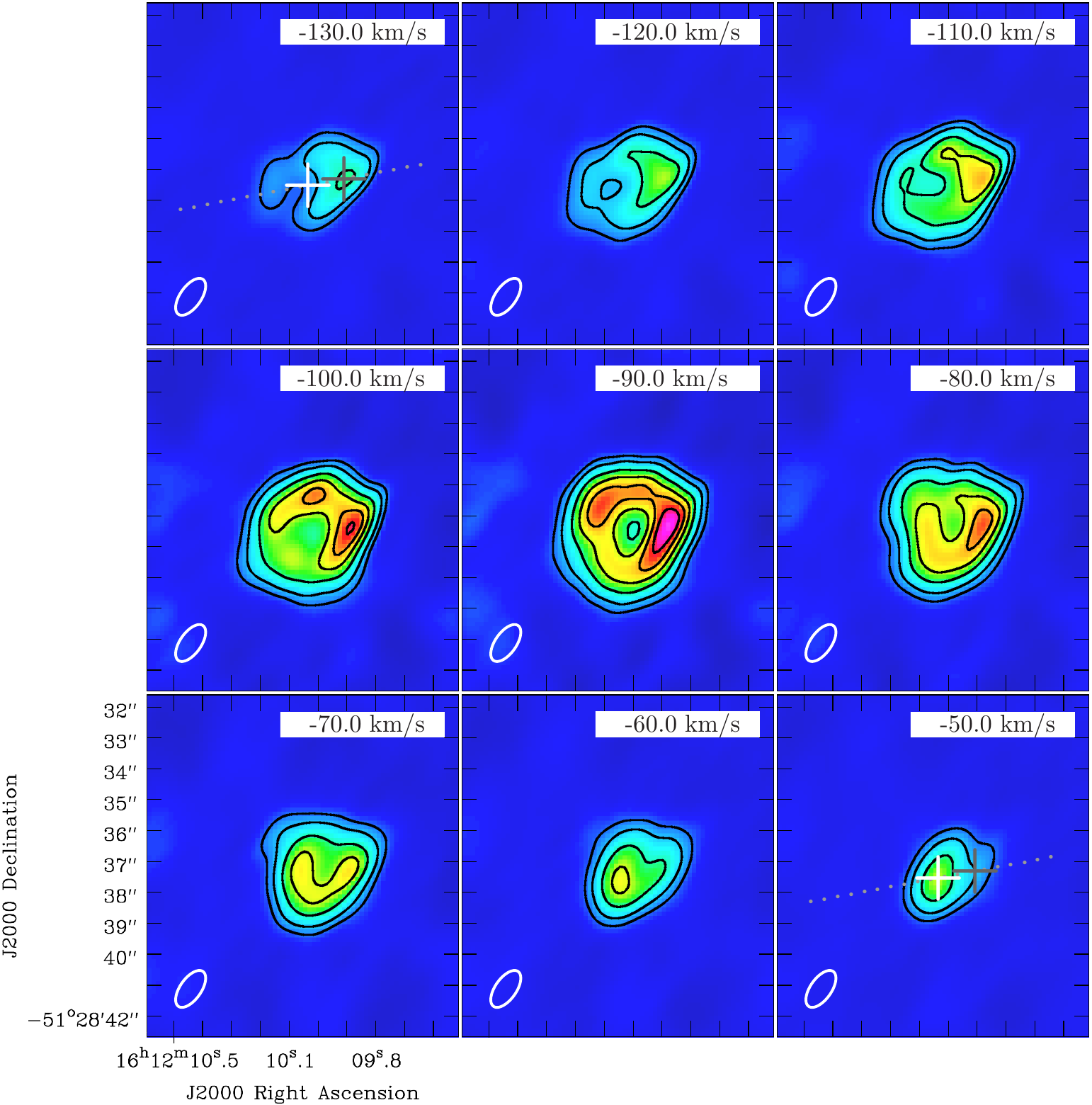}
	\end{center}
	\caption{Channel maps of the SiO(8-7) line emission. The contours are at 3, 6,
12, 18, 24 and 30 times the rms defined for the SiO emission (0.14 Jy
beam$^{-1}$). The velocity of each channel is indicated in the upper right
corner of each box. The channels are separated by 10 \kms. In the central
channel map, toward the systemic velocity of the source ($-88.9$ \kms), the
emission shows a clear ring-type or projected shell structure, with the
inner hole well determined at the 18 $\sigma$ contour. The ALMA synthesized beam
is shown at the bottom left corner in each box. The gray and white crosses define the positions considered in the blue and red lobes to set the symmetry axis of the outflow, represented by the gray dotted line.}
	\label{fig:channel_map}
\end{figure}

\begin{figure}[h]
	\begin{center}
	\includegraphics[width=0.6\textwidth]{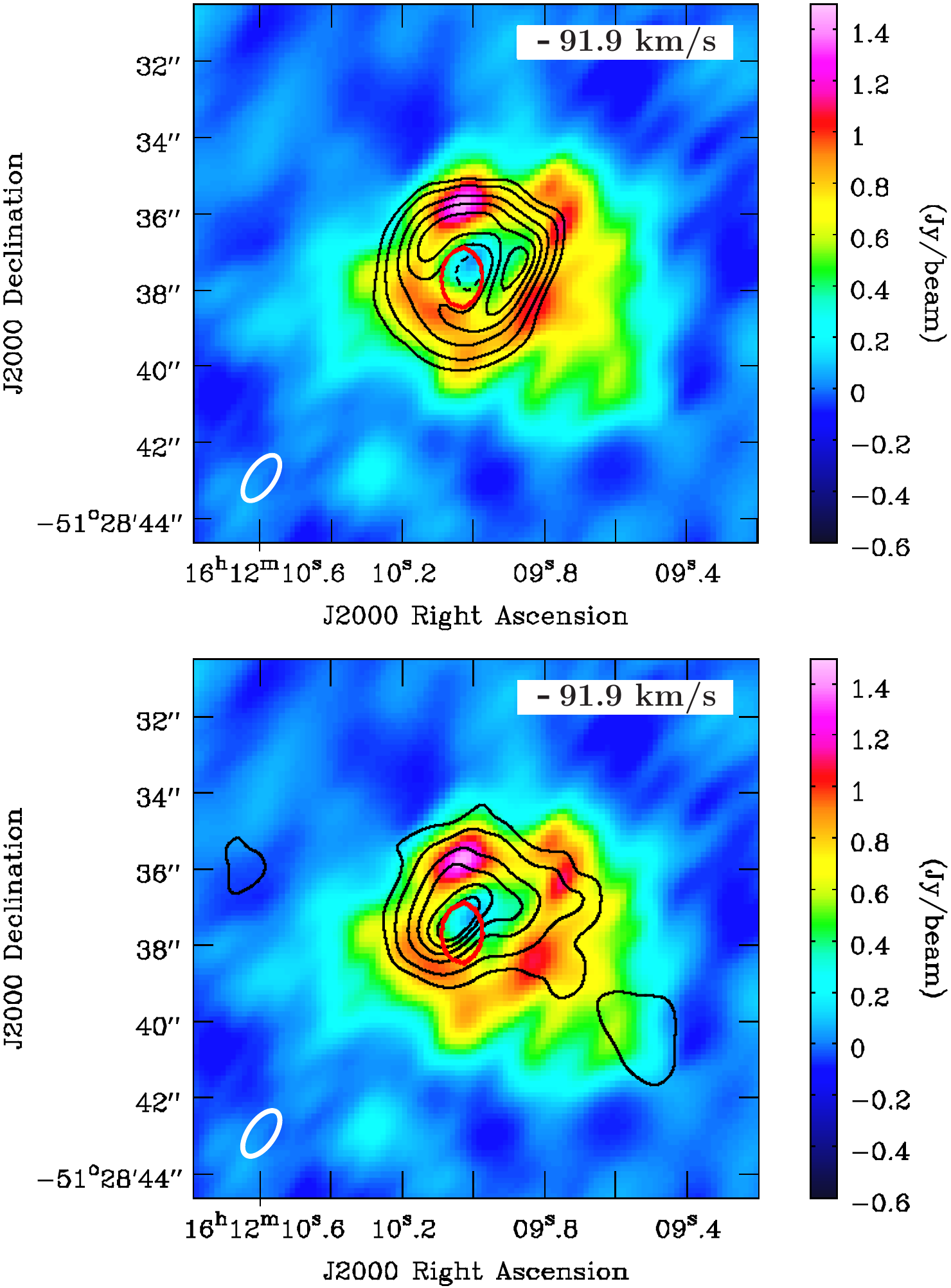}
	\end{center}
	\caption{$Top:$ \hcopt\ (4-3) emission (in colors) 
of the \gtresoutflow\ source
at a velocity of $-91.9$ \kms, where the peak of flux density is found. The
overlaid contours in black correspond to the SiO(8-7) emission at the same
velocity, at 3, 6, 12, 18, 24 and 30 $\sigma$ (with $\sigma$=0.14 Jy beam$^{-1}$). The dashed contour represent the 12 $\sigma$ emission toward the cavity observed in SiO, which is coincident spatially with the ones observed with
\hcopt. The red contour shows the 50\% of emission (peak of 158 mJy beam$^{-1}$) of the radio source detected at 8.6 GHz toward this source, presented by Merello et al. (2013). The peak position of the radio continuum
observations is inside the cavity described by the SiO and \hcopt\ 
emissions. $Bottom:$ \hcopt\ (4-3) emission (in colors), with overlaid black contours of the 0.86 mm continuum emission obtained with ALMA. The contours are at 3, 6, 12, 18, 24, 30 $\sigma$ (with $\sigma$=0.02 Jy beam$^{-1}$). The continuum emission peaks, within errors, in the center of the cavity shown in the \hcopt\ emission.  }
	\label{fig:h13cop_sio}
\end{figure}

\begin{figure}[h]
	\begin{center}
		\includegraphics[angle=0, width=0.4\textwidth]{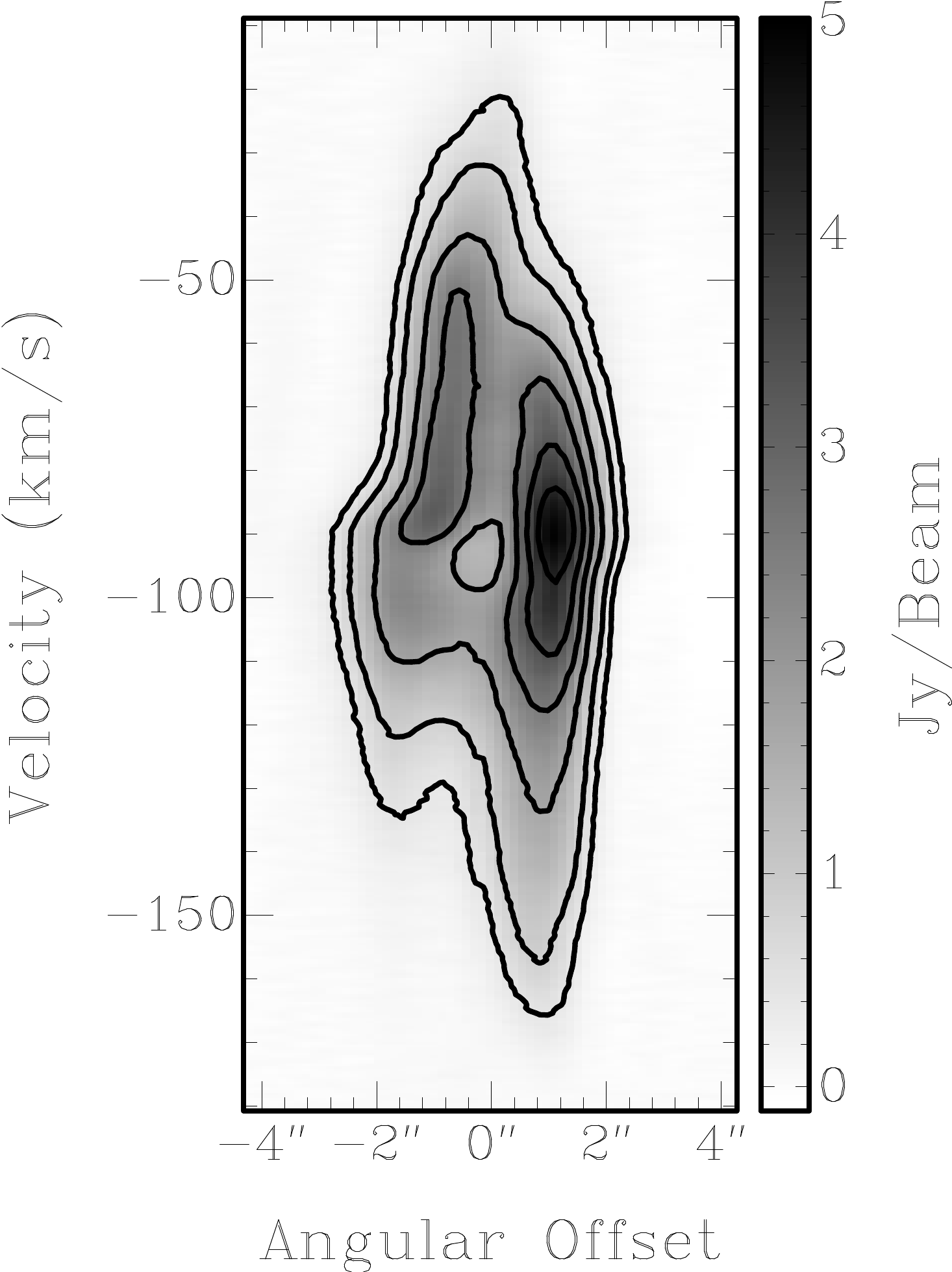}\ \ 
		\includegraphics[angle=0, width=0.4\textwidth]{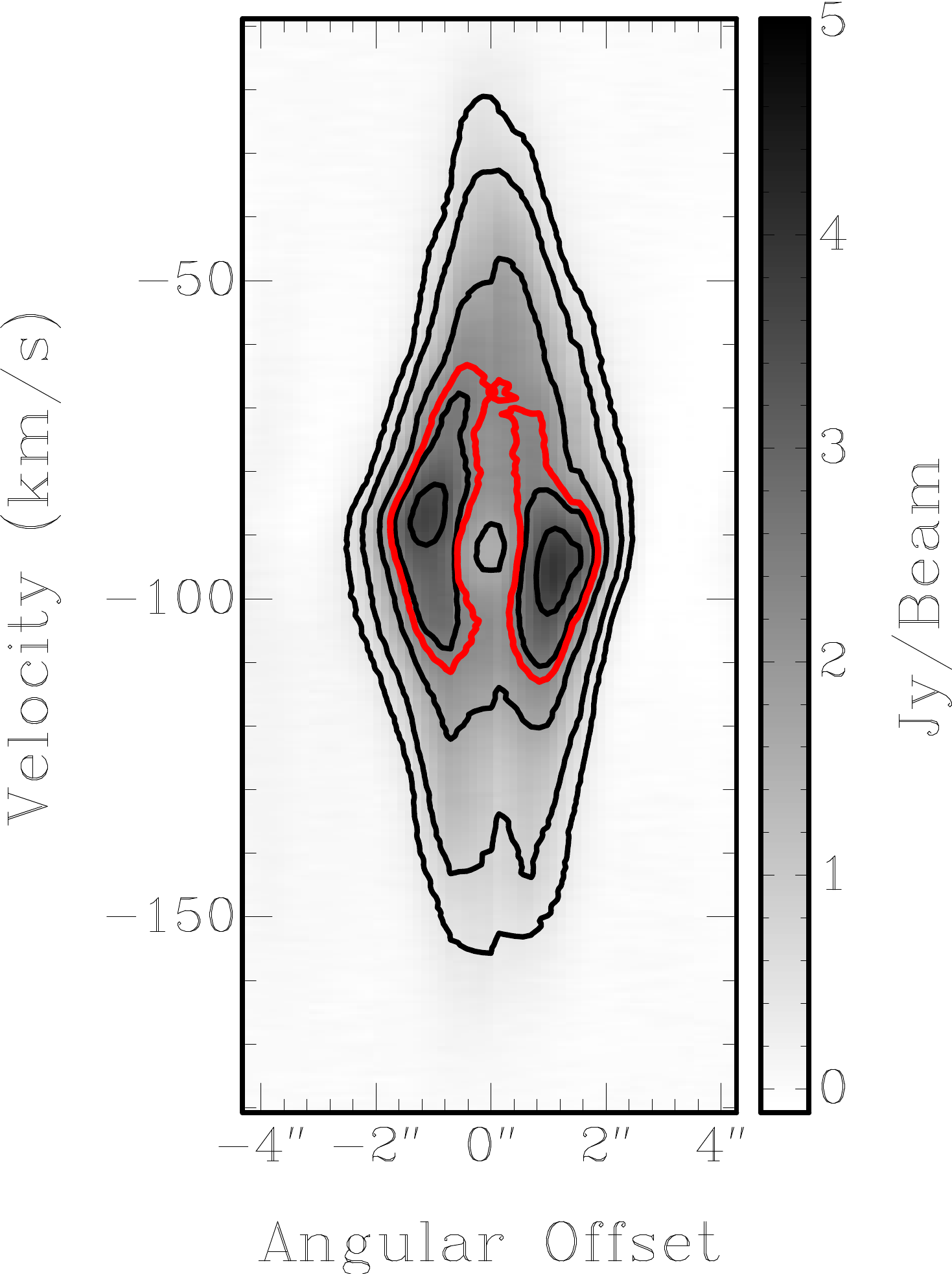}
		
		\includegraphics[angle=0, width=0.44\textwidth]{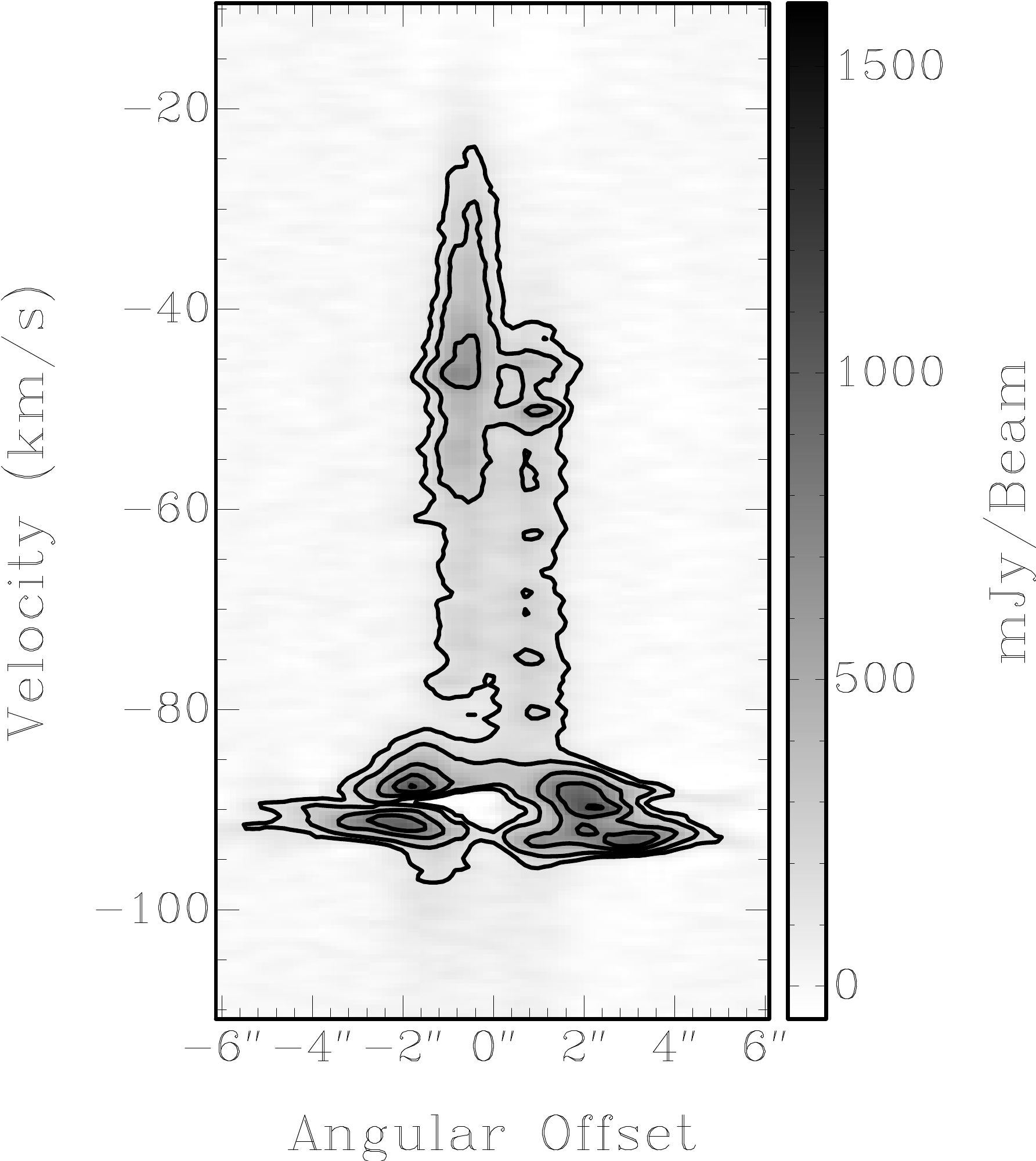}\ \ 
		\includegraphics[angle=0, width=0.44\textwidth]{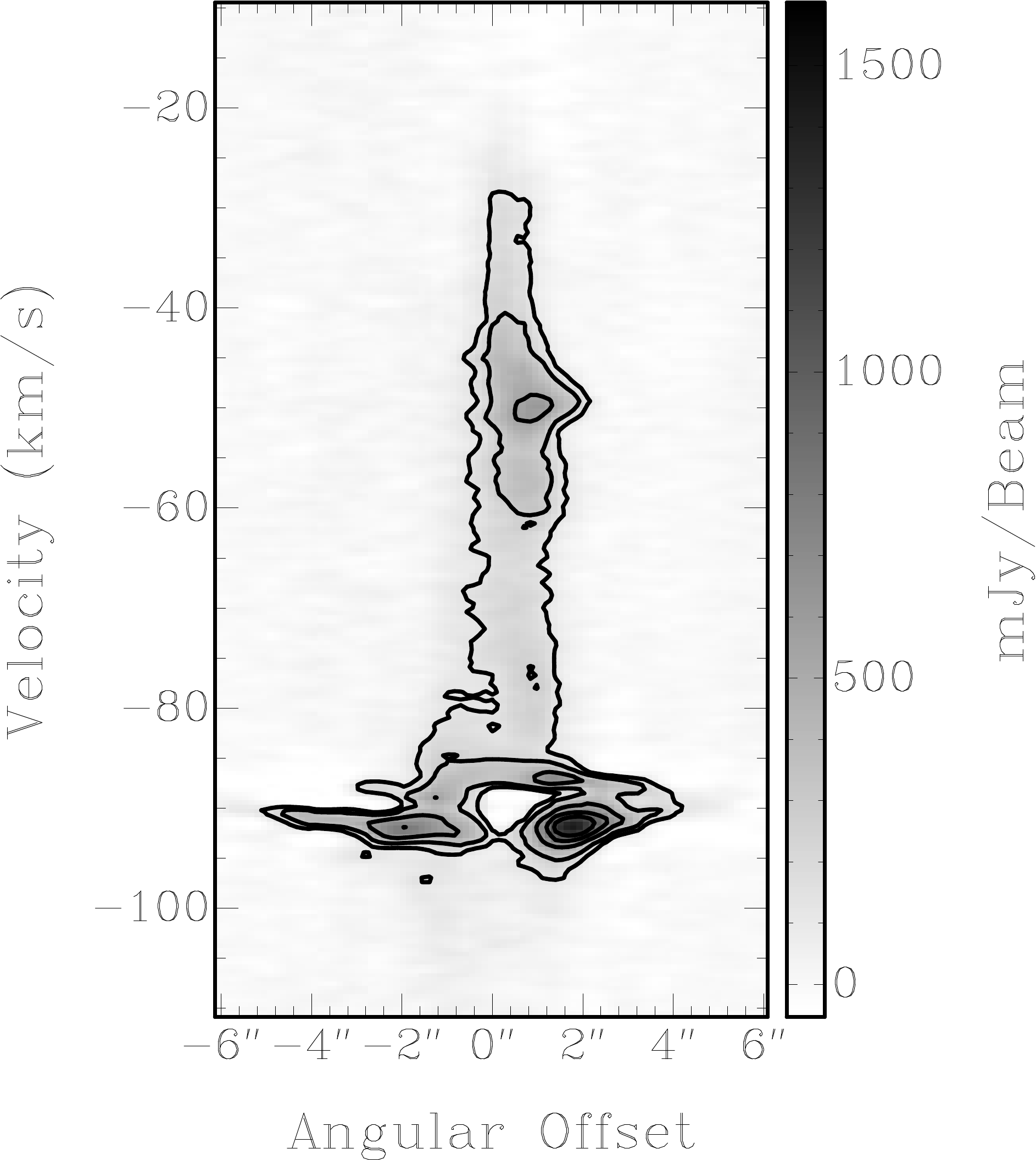}
	\end{center}
	\caption{Position-velocity diagrams of the SiO(8-7) ($top$) and H$^{13}$CO$^+$(4-3) ($bottom$)
emission of the \gtresoutflow\ source. The slices are centered at the cavity
observed in the SiO map ($\alpha_{2000} = 16^h12^m09.985^s$, $\delta_{2000} =
-51\arcdeg28\arcmin 37.42\arcsec$). 
Slices along the axis of symmetry of the outflow (primary axis),
and perpendicular to this (secondary axis), are displayed in gray scale, with
3, 6, 12, 18, 24 and 30 $\sigma$ contours, with a slit of 4.5\arcsec\ for the
SiO emission, and 6\arcsec\ for \hcopt. 
$Upper\ left:$ Slice of the SiO emission,
running along the axis of symmetry (position angle of
102.5\arcdeg). $Upper\ right:$ Slice of the SiO emission, running perpendicular
to the axis of symmetry (position angle of 12.5\arcdeg). An additional 16 $\sigma$ contour in red is shown, extending between -111.9 and -63.8 \kms. $Bottom\ left:$ Slice of the \hcopt\ emission, running along the axis of symmetry. $Bottom\ right:$ Slice of the \hcopt\ emission, running
perpendicular to the axis of symmetry.}
	\label{fig:pvmap1}
\end{figure}

\end{document}